\documentclass[aps,prb,groupedaddress,amsmath,twocolumn,amssymb,titlepage]{revtex4-1}
\usepackage{hyperref}
\usepackage{graphicx}
\usepackage{enumerate}
\usepackage{bm,bbm,dsfont}

\begin{document}


\title{Time Dependent Variational Principle with Ancillary Krylov Subspace}


\author{Mingru Yang}
\email{mingruy@uci.edu}
\affiliation{Department of Physics and Astronomy, University of California, Irvine, CA 92697, USA}
\author{Steven R. White}
\affiliation{Department of Physics and Astronomy, University of California, Irvine, CA 92697, USA}


\date{\today}

\begin{abstract}
We propose an improved scheme to do the time dependent variational principle (TDVP) in finite matrix product states (MPS) for two-dimensional systems or one-dimensional systems with long range interactions. We present a method to represent the time-evolving state in a MPS with its basis enriched by state-averaging with global Krylov vectors. We show that the projection error is significantly reduced so that precise time evolution can still be obtained even if a larger time step is used. Combined with the one-site TDVP, our approach provides a way to dynamically increase the bond dimension while still preserving unitarity for real time evolution. Our method can be more accurate and exhibit slower bond dimension growth than the conventional two-site TDVP.
\end{abstract}


\maketitle


At the heart of the success of the density matrix renormalization group (DMRG)\cite{PhysRevLett.69.2863,PhysRevB.48.10345} for approximating the ground states of one- and two-dimensional lattice systems is the matrix product state (MPS) representation underlying it. MPS are ideal for one-dimensional gapped ground states\cite{PhysRevB.73.094423,Hastings_2007}, and are also a powerful approximation for one-dimensional gapless ground states and ground states of finite-width two-dimensional\cite{doi:10.1146/annurev-conmatphys-020911-125018} cylinders and strips.

MPS are also useful in solving the {\it time-dependent} Schr{\"o}dinger equation. Vidal's time-evolving block decimation (TEBD) method\cite{PhysRevLett.91.147902,PhysRevLett.93.040502,PhysRevLett.98.070201} can be framed as a slight change to the DMRG sweeping algorithm, called the time-dependent DMRG method (tDMRG)\cite{PhysRevLett.93.076401}. Since the invention of TEBD, a number of variations have been developed, and have been widely used, for example in computing spectral functions\cite{PhysRevB.85.205119} and in simulating the dynamics of cold atom systems\cite{Trotzky2012}.

Treating the time evolution of systems with long-range interactions is more difficult. In its original form, TEBD handles only nearest-neighbor interactions. A simple modification, exploiting swap gates to move sites which are not next to each other to be temporarily adjacent, is effective for systems with only a modest number of beyond-nearest-neighbor interactions. Alternatively, an approach\cite{PhysRevB.91.165112} to approximate the time evolution operator in terms of a matrix product operator (MPO) was developed to treat long-range interactions efficiently, and has proven successful in calculating the response function of the spin-1/2 Haldane-Shastry model\cite{PhysRevB.91.165112}. Another approach\cite{PhysRevB.72.020404} allowing the simulation of dynamics with long-range interactions is based on Runge-Kutta, and an improved version of it was recently found useful in treating the dynamics of chemical systems\cite{doi:10.1021/acs.jctc.7b00682}.

\begin{figure}
\includegraphics[width=8.5cm]{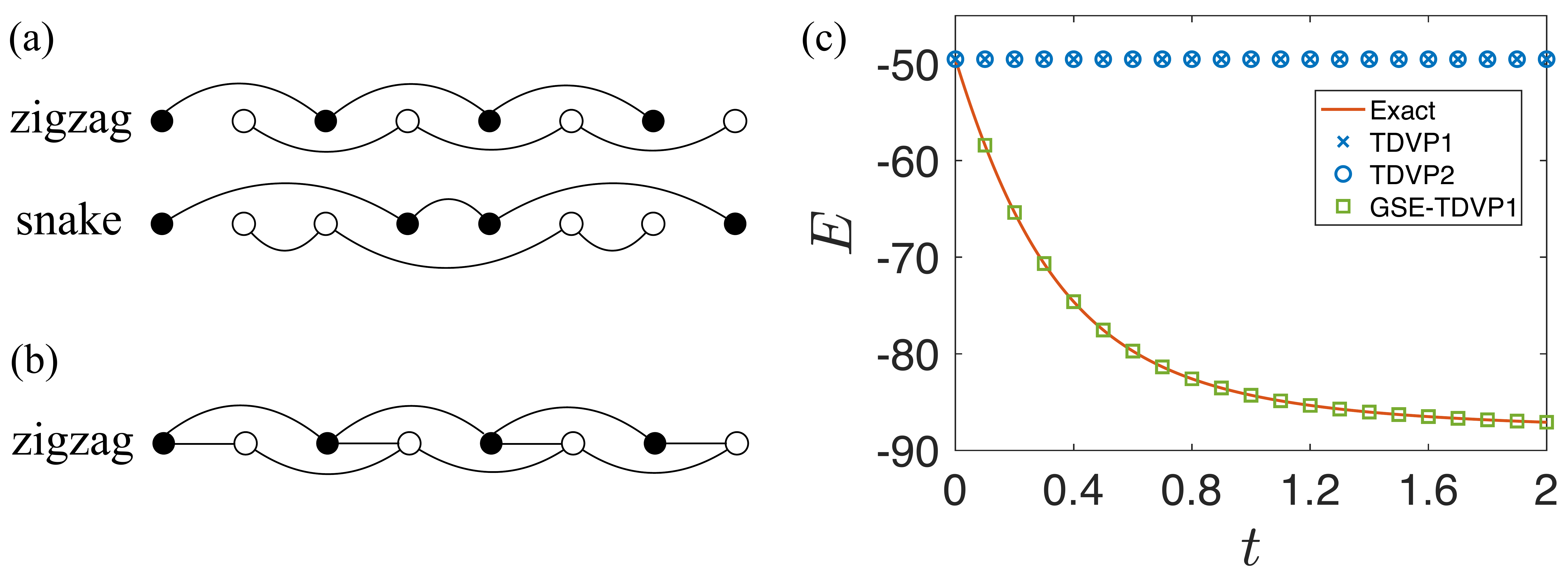}
\caption{\label{fig:tba} (a) Interactions in the rung-decoupled Heisenberg ladder after mapping the lattice to a one-dimensional geometry by using zigzag or snake path. Black(white) circles are sites in the first(second) leg. Only circles connected by lines have interactions between them. (b) Interactions in the rung-coupled Heisenberg ladder. (c) Failure of the one-site TDVP (TDVP1) and two-site TDVP (TDVP2) for the imaginary time evolution of the rung-decoupled Heisenberg ladder of leg length 100. GSE-TDVP1 is our method. The zigzag path is used.}
\end{figure}

Some of the most attractive MPS time evolution methods which can deal with long-range interactions are based on the time dependent variational principle (TDVP)\cite{dirac_1930,PhysRevLett.107.070601}, where the time-dependent Schr{\"o}dinger equation is projected to the tangent space of the MPS manifold of fixed bond dimension at the current time. The manifold of all possible MPS's with a particular bond dimension can be thought of as a constraining surface for the time evolution, which in the case of real time can be thought of analogously to classical motion under a constraint. As in classical mechanics, which possesses a symplectic structure, the probability and energy is automatically conserved, and so are the other integrals of motion provided that the corresponding symmetry transformation does not take the state out of the manifold\cite{PhysRevB.88.075133}. Energy and probability are not conserved in most other time dependent MPS methods due to the necessity of truncation to keep an efficient MPS representation.

The initial implementation\cite{PhysRevLett.107.070601} of TDVP simultaneously updates all MPS tensors and suffers from numerical instability. The difficulties were overcome later by an alternative integration scheme\cite{PhysRevB.94.165116} based on a Lie-Trotter decomposition of the tangent space projector. This algorithm, in the imaginary time case, as the time step goes to infinity, is equivalent to DMRG. The integration scheme, like DMRG, can be based either on one site or two sites. In the one-site method, the symplectic property is retained, but the bond dimension of the MPS cannot increase to accommodate increased entanglement. The two-site scheme involves a truncation process that allows evolution to a manifold with higher or lower bond dimension, although the symplectic property is lost.

A highly desirable feature of TEBD methods is that, aside from well-understood Trotter decomposition errors, the errors in the evolution all stem from the singular value decomposition (SVD) truncation of an MPS bond, which is precisely quantified and controlled. This is a much better situation than in ground-state DMRG, where the search for the ground state may get stuck in a local 
minimum\footnote{\label{ft:local}In DMRG, we can alleviate the problem using density matrix perturbation\cite{PhysRevB.72.180403}, subspace expansion\cite{PhysRevB.91.155115}, more center sites\cite{PhysRevA.99.022509}, or swap gates.}\cite{PhysRevB.72.180403,PhysRevB.91.155115,PhysRevA.99.022509}.

Unfortunately, TDVP methods can fail in an uncontrolled manner, in some ways similar to DMRG. This failure in a simple situation is illustrated in Fig. \ref{fig:tba}(c). The system is a rung-decoupled Heisenberg ladder--when starting the time evolution in a product state, both the one- and two-site schemes fail to time-evolve at all.

This example illustrates some of the significant projection errors that can arise in TDVP. In the one-site scheme, TDVP evolution of a product state stays in a product state for any Hamiltonian. In addition, after being mapped to a one-dimensional geometry by using the zigzag (or snake) path, there exist two subsets of sites that are disconnected by interactions (Fig. \ref{fig:tba}(a)). The two sites involved in the local update at each bond (or each odd bond for the snake path) belong to the two disconnected subsets respectively, so two-site TDVP fails to build up the intra-leg entanglement unless the tangent space of the MPS manifold already contains the necessary degrees of freedom. When the initial state is a product state, i.e. an MPS of bond dimension one, the limitations of the tangent space are the most severe. In a more typical system, e.g. with the rung coupling turned on, as shown in Fig. \ref{fig:tba}(b), no disconnected subsets exist--although the two sites at even bonds are not directly connected by nearest-neighbor interactions, they are connected through a remote site. Therefore, in such situations the errors of two-site TDVP can be reduced by using a smaller time step.

There have been some tricks to enlarge the bond dimension of the product state to be time evolved by the one-site TDVP. The original MPS can be embedded in a MPS with larger bond dimension by filling up zeros\cite{PhysRevB.100.104303,kloss2020studying}, but this approach does not help to immediately reduce the projection errors\cite{PhysRevB.88.075133}. Another way is to use several DMRG sweeps to introduce some noise which artificially increases the bond dimension\cite{PAECKEL2019167998}. But our test shows that for two-dimensional systems and long-range interactions, large errors emerge after a short time even though the bond dimension has been enlarged substantially. Increasing the bond dimension to a large value at the beginning and keeping it through the whole time evolution is in fact very inefficient, since the bond dimension needed for the initial time might otherwise be much smaller than at later times. So it is suggested\cite{PhysRevB.99.054307,PAECKEL2019167998} to first use two-site TDVP to increase the bond dimension for the initial sweeps and then switch to the one-site TDVP. However, in the example in Fig. \ref{fig:tba}(c), two-site TDVP also fails.

In this paper we provide an effective way to fix the projection errors in TDVP methods, which works for both the extreme case above and other systems with long-range interactions even if a large time step is used. We show how one can expand the MPS manifold thereby enlarging the tangent space before a TDVP time step, to make it contain the true direction of motion. This enlargement is based on a subspace expansion\cite{doi:10.1137/140953289,PhysRevB.91.155115} by global Krylov vectors. Unlike in the exact diagonalization\cite{10.2307/2158085,10.1145/285861.285868}, in our algorithm the global Krylov vectors serve as ancillary MPS's to enrich the basis of the time-evolving MPS through the gauge degree of freedom, thus avoiding the problems of loss of orthogonality and production of unnecessarily highly entangled state\cite{PAECKEL2019167998}. Our method has a modest computational cost compared to that of a TDVP time step. Also, it is easy to turn off the tangent space enlargement in regimes where ordinary TDVP methods work well.

\section{Algorithms}
\label{sec:1}
\begin{figure}
\includegraphics[width=5.3cm]{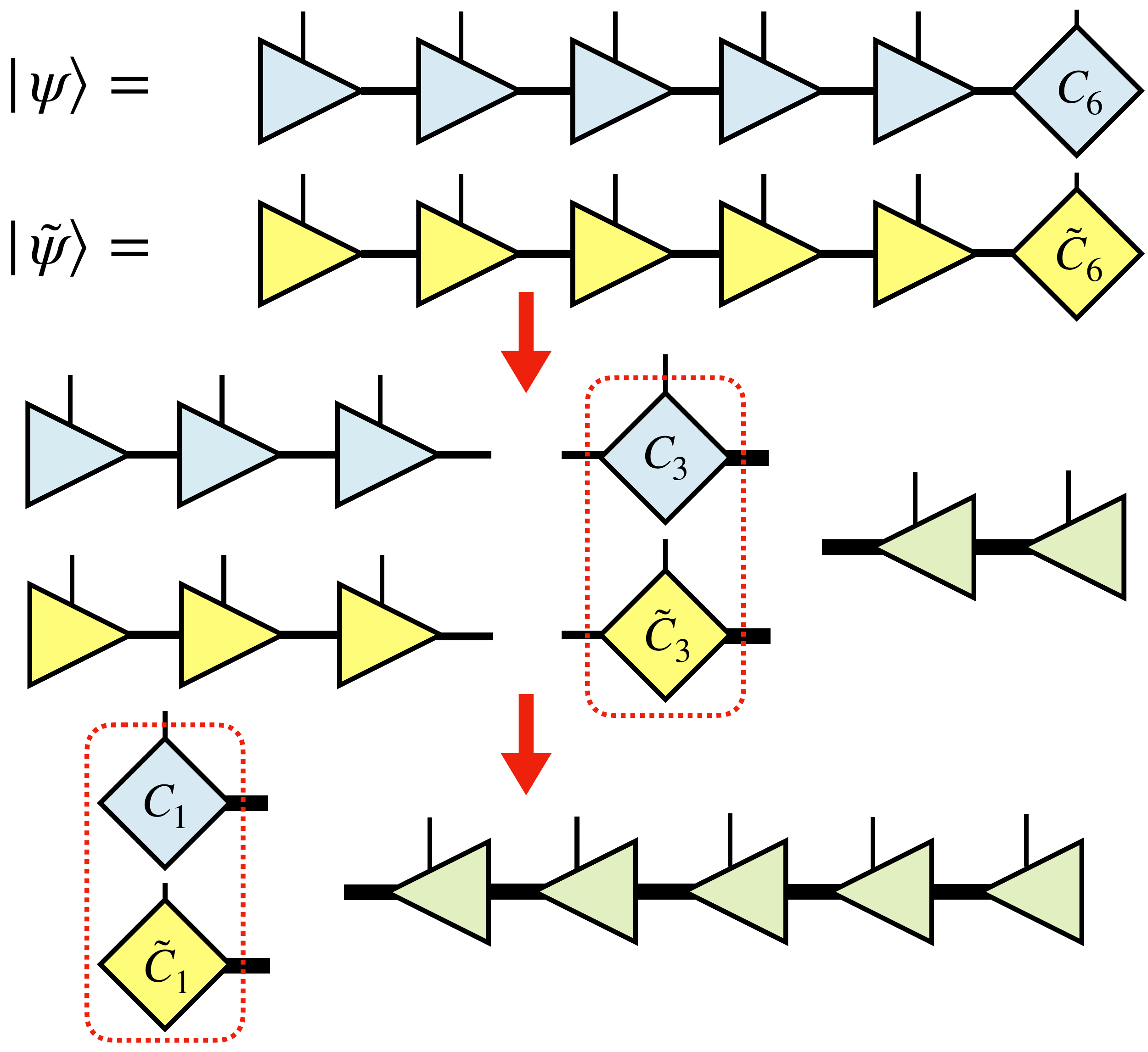}
\caption{\label{fig:3} Basis extension of $|\psi\rangle$ by $|\tilde{\psi}\rangle$. The diamond $C_i$($\tilde{C}_i$) is the orthogonality center of $|\psi\rangle$($|\tilde{\psi}\rangle$) at site $i$. The triangle pointing to the right(left) is left(right)-orthonormal. The red dashed rounded rectangle means forming a direct sum of the tensors inside it. The bolder bond has the larger bond dimension. }
\end{figure}
Normally, in working with MPS, we always seek the smallest MPS to represent a particular state. Here, we find that it is very useful to temporarily create an inefficient MPS representation of the current time-evolved state by expanding the MPS to represent both the current state and a short Krylov expansion of it. The Krylov expansion is generated with a standard MPS algorithm\cite{denmatapply,PAECKEL2019167998} which is capable of producing nonlocal entanglement. With the expanded manifold coming from this inefficient representation, a subsequent TDVP time step is accurate and reliable.

In this section, we first introduce the basis extension of an MPS by another one and discuss the trick to exactly preserve the information of the original time-evolved state. Then we discuss several issues of generating the global Krylov vectors. The algorithm is summarized in section \ref{subsec:sum}.

\subsection{Basis extension}
\label{subsec:alg}
The MPS representation of a physical state is not unique. We can utilize the gauge degrees of freedom to extend the basis at each bond so as to get an MPS with enlarged bond dimension without changing the physical state. Specifically, this property is used in our method to yield an MPS with its basis extended by other MPS's, which is reminiscent of the multi-state targeting\cite{PhysRevB.85.205119,Wall_2012} approach frequently used in the early days of DMRG to deal with excited states simultaneously with the ground state. In the old DMRG language, targetting more than one state (also called state-averaging) is like having a mixed state, and one averages density matrices at each DMRG step. In the modern MPS language, the state-averaging is done by creating an extra index to label the states involved. In the description of our algorithm, we incorporate both formulations, which are equivalent.

Suppose we have an MPS of a state $|\psi\rangle$, and we wish to extend the MPS bond basis of $|\psi\rangle$ by that of another state $|\tilde{\psi}\rangle$. Suppose both states are in  left-canonical form (Fig. \ref{fig:3}),
\begin{equation}
|\psi\rangle= \sum_{s_1\cdots s_N}A_1^{s_1}\cdots A_{N-1}^{s_{N-1}}C_N^{s_N}|s_1\cdots s_N\rangle,
\end{equation}
and similarly for $|\tilde{\psi}\rangle$. Here $A_1^{s_1}$ is a $1\times m_1$ matrix and $C_N^{s_N}$ is a $m_{N-1}\times 1$ matrix, where $m_i$ is the bond dimension between site $i$ and $i+1$. First we write the direct sum formally,
\begin{equation}
\label{eq:drctsum}
\begin{aligned}
\left[\begin{array}{cc}
|\psi\rangle\\
|\tilde{\psi}\rangle
\end{array}\right]
=\sum_{s_1\cdots s_N}A_1^{\prime s_1}\cdots A_{N-1}^{\prime s_{N-1}}C_N^{\prime s_N}|s_1\cdots s_N\rangle\\
=\sum_{s_1\cdots s_N}
\left[\begin{array}{cc}
A_1^{s_1} & 0\\
0 & \tilde{A}_1^{s_1}
\end{array}\right]
\cdots
\left[\begin{array}{cc}
A_{N-1}^{s_{N-1}} & 0\\
0 & \tilde{A}_{N-1}^{s_{N-1}}
\end{array}\right]
\left[\begin{array}{cc}
C_{N}^{s_{N}}\\
\tilde{C}_{N}^{s_{N}}
\end{array}\right]\\
|s_1\cdots s_N\rangle,
\end{aligned}
\end{equation}
where the second line defines the primed matrices in the first line. Now $A_1^{\prime s_1}$ is $2 \times (m_1+\tilde{m}_1)$, and the product of matrices gives a $2\times 1$ coefficient matrix (for each set of $s_1\cdots s_N$). The extra two-dimensional index, attached to the first site, picks either $|\psi\rangle$ or $|\tilde{\psi}\rangle$. We can now compress the expanded MPS by doing SVD with truncation iteratively from the right end to the left. At site $i$, we perform $C_i^{\prime s_i}=U_i^\prime S_i^\prime B_i^{\prime s_i}$. We continue at the next site $i-1$ with
\begin{equation}
\label{eqn:svd}
C_{i-1}^{\prime}=
\left[\begin{array}{cc}
C_{i-1}\\
\tilde{C}_{i-1}
\end{array}\right]
=
\left[\begin{array}{cc}
A_{i-1}C_iB_i^{\prime\dag}\\
\tilde{A}_{i-1}\tilde{C}_iB_i^{\prime\dag}
\end{array}\right],
\end{equation}
where the local physical indices have been omitted. Note that it is unnecessary to explicitly implement the form in Eq. (\ref{eq:drctsum}). Because of the block diagonal form of $A_{i-1}^{\prime}$, it will be more efficient to move the orthogonality center to site $i-1$ separately for $|\psi\rangle$ and $|\tilde{\psi}\rangle$ as in Eq. (\ref{eqn:svd}) than simply doing $C_{i-1}^{\prime}=A_{i-1}^{\prime}U_i^{\prime}S_i^{\prime}$. Eventually we will end up with
\begin{equation}
\label{eqn:common}
\left[\begin{array}{cc}
|\psi\rangle\\
|\tilde{\psi}\rangle
\end{array}\right]
=\sum_{s_1\cdots s_N}
\left[\begin{array}{cc}
C_1^{s_1}\\
\tilde{C}_1^{s_1}
\end{array}\right]
B_2^{\prime s_2}\cdots B_N^{\prime s_N}
|s_1\cdots s_N\rangle.
\end{equation}
Note that this common representation for both states can be used to sum them, with arbitrary coefficients, $a|\psi\rangle+b|\tilde{\psi}\rangle$, by performing the same operation on $C_1^{s_1}$ and $\tilde{C}_1^{s_1}$. (One would then want to reorthogonalize from left to right, if one were only interested in the sum.) In our case, we want only $|\psi\rangle$,
so we simply throw out $\tilde{C}_1^{s_1}$, which gives us a right-canonical MPS of $|\psi\rangle$ with its bond basis extended by $|\tilde{\psi}\rangle$ and its orthogonality center $C_1$ not full column rank. 

However, the algorithm in this form has a drawback--it is not convenient to treat $|\tilde{\psi}\rangle$ less accurately than $|\psi\rangle$, or, in particular, to retain $|\psi\rangle$ exactly. To solve this problem, at each site $i$, instead of simply SVDing and truncating $C_i^{\prime}$, we select the most important basis from $|\tilde{\psi}\rangle$ and orthogonalize them against the existing basis of $|\psi\rangle$ before combining them.

In the following we explain the reformed algorithm in more generic settings, with more than just two states---suppose we have $k$ MPS's now and want to use the latter $k-1$ MPS's, $|\tilde{\psi}_l\rangle$, to extend the basis of the first one, $|\psi\rangle$. The extra time complexity to deal with $k$ MPS's will be larger by $O(k)$. At site $i$ (suppressing the index $i$ below), we first SVD the site tensor of $|\psi\rangle$ without truncation, i.e. $C=USB$, and form the null-space projection operator $P=1-B^\dagger B$. We sum the reduced density matrices of the other $k-1$ MPS's
\begin{equation}
\tilde{\rho}=\sum_{l=1}^{k-1}\tilde{\rho}_l,
\end{equation}
where $\tilde{\rho}_l=\tilde{C}_l^\dagger\tilde{C}_l$. If $P\neq 0$, we then project $\tilde{\rho}$ by $P$,
\begin{equation}
\label{eqn:mix}
\bar{\rho}\equiv P\tilde{\rho}P.
\end{equation}
Diagonalizing and truncating, $\bar{\rho}=\bar{B}^{\dagger}\bar{S}^2\bar{B}$. (This is equivalent to projecting each $\tilde{C}_l$ by $P$ and SVDing the direct sum of them.) The rows of $\bar{B}$ are orthogonal to those of $B$, i.e. $\bar{B} B^\dagger=0$, so we can enlarge the row space of $B$ by the direct sum
\begin{equation}
B'=
\left[\begin{array}{cc}
B\\
\bar{B}
\end{array}\right],
\end{equation}
forming the new right-orthonormal MPS tensor at site $i$ in Eq. (\ref{eqn:common}).

\subsection{Krylov subspace}
What states do we use to enlarge the basis for $|\psi\rangle$? It is natural to consider the time evolution which TDVP is implementing.  Consider the wavefunction
evolved for a short time: 
\begin{equation}
\label{eqn:taylor}
|\psi(t+\Delta t)\rangle=\exp{(-\mathrm{i}\hat{H}\Delta t)}|\psi(t)\rangle
\approx\sum_{l=0}^{k-1}\frac{(-i\Delta t)^l}{l!}\hat{H}^l|\psi(t)\rangle,
\end{equation}
where $t+\Delta t$ can be either imaginary or real. A problem with utilizing Eq. (\ref{eqn:taylor}) directly for the time evolution is that the expression converges slowly in $k$. Instead, we use only a few of the terms appearing in Eq. (\ref{eqn:taylor}) to extend the basis set of $|\psi(t)\rangle$ before time evolution by a following TDVP sweep. Specifically, in the algorithm presented here we consider an MPS extended to represent the Krylov subspace of order $k$
\begin{equation}
\mathcal{K}_k(\hat{H},|\psi\rangle)=\mathrm{span}\{|\psi\rangle,\hat{H}|\psi\rangle\dots, \hat{H}^{k-1}|\psi\rangle\},
\end{equation}
where the order $k$ of the Krylov subspace is quite small.

There are three technical issues which need further elaboration. First, since the norm of $\hat{H}^l|\psi(t)\rangle$ grows exponentially with $l$, for numerical stability, we either normalize each MPS separately, or, motivated by the first-order expansion of the time evolution operator, replace them by
\begin{equation}
\label{eqn:set}
(1-\mathrm{i}\tau\hat{H})|\psi(t)\rangle,\dots,(1-\mathrm{i}\tau\hat{H})^{k-1}|\psi(t)\rangle,
\end{equation}
where $\tau$ is a small parameter to be tuned to make sure the norm of $(1-\mathrm{i}\tau\hat{H})^l|\psi(t)\rangle$ do not blow up. For imaginary time evolution, we can choose $\mathrm{i}\tau$ to be $\lambda^{-1}$, where $\lambda$ is approximately the highest energy of the excited states. For real time evolution, the choice of $\tau$ does not  matter as much and we can simply set $\tau=\Delta t$. Note the states in Eq. (\ref{eqn:set}) still span the Krylov subspace.

The second issue is, how do we apply $\hat{H}$ efficiently? When the bond dimension of the MPO of $\hat{H}$ is small, we can use the density matrix approach\cite{denmatapply} (which is exact); otherwise we can use the variational approach\cite{PAECKEL2019167998}. The complexity of applying $\hat{H}$ at each site is comparable to one Lanczos iteration  used to integrate the local effective equations at a site in TDVP. Usually the number of iterations needed at a site in TDVP is much larger than $k$, so the time cost of the application of $\hat{H}$ is subleading.

The third issue is, how do we control the bond dimension of $\hat{H}^l|\psi(t)\rangle$, which grows fast with increasing $l$? Fortunately, we have found that for a reasonable choice of time step size, $k=3$ can already provide good accuracy. Furthermore, we do not need as high an accuracy in  $\hat{H}^l|\psi(t)\rangle$ as compared to $|\psi(t)\rangle$ itself. Therefore we utilize a considerable truncation cutoff $\epsilon_K$ in applying $\hat{H}^l$.

There are a few other places in our algorithm where truncation is necessary. When diagonalizing the projected {\it{mixing}} of reduced density matrices in Eq. (\ref{eqn:mix}), we use a truncation cutoff $\epsilon_M$. This truncation controls the number of states added into $|\psi (t)\rangle$. We also define a truncation cutoff $\epsilon$ which is applied during the TDVP time step. This cutoff is only used when we do not require the time evolution to be exactly unitary, such as in imaginary time evolution.

\subsection{Subspace expansion}
\label{subsec:sum}
We now describe our {\it{global subspace expansion}} (GSE) TDVP algorithm. Because of the Krylov subspace expansion, we do not need to use a two-site method. Combined with the one-site TDVP, one time step of our algorithm (GSE-TDVP1) consists of
\begin{enumerate}
\item Construct the MPO for $1-\mathrm{i}\tau\hat{H}$, and apply it $k-1$ times to $|\psi (t)\rangle$ and get the set of MPS's in Eq. (\ref{eqn:set}). (Controlled by $\epsilon_K$.)
\item Do basis extension for $|\psi (t)\rangle$ as described in \ref{subsec:alg}, ending with the orthogonality center at the first site. (Controlled by $\epsilon_M$.)
\item From left to right, do a conventional one-site TDVP sweep, and then sweep from right to left. (Controlled by $\epsilon$.)
\end{enumerate}

Let the bond dimension of $|\psi(t)\rangle$ before step 1 be $m$ and the bond dimension of the MPO for $1-\mathrm{i}\tau\hat{H}$ be $w$. Usually we choose $\epsilon_K$ to make the bond dimension of each of the Krylov vectors in Eq. (\ref{eqn:set}) no bigger than $m$. Then, if we apply the MPO variationally\cite{PAECKEL2019167998}, the complexity of step 1 is $O(km^3wd)$. Let the bond dimension of $|\psi(t)\rangle$ after step 2 be $m'$. Then the complexity of step 2 is $O(k{m'}^3d^2+{m'}^3d^3)$. The one-site TDVP has a complexity $O(l{m'}^3wd)$, where $l$ is the number of Lanczos steps used at each local update. As described in the last section, usually $l\gg k$ so the cost of step 1 and 2 will be comparable to or less than step 3. In our benchmarks for real time evolution, we find that if we use $k=5$ and get $m'\approx 3m$, the time taken for step 1 and 2 is about 1/3 of that for step 3; given the same bond dimension $m'$, the time taken for one GSE-TDVP1 time step is about 36\% of that for a conventional two-site TDVP step.

\section{Benchmarks}
\label{sec:2}
\subsection{Imaginary time evolution}
\begin{figure}
\includegraphics[width=8.4cm]{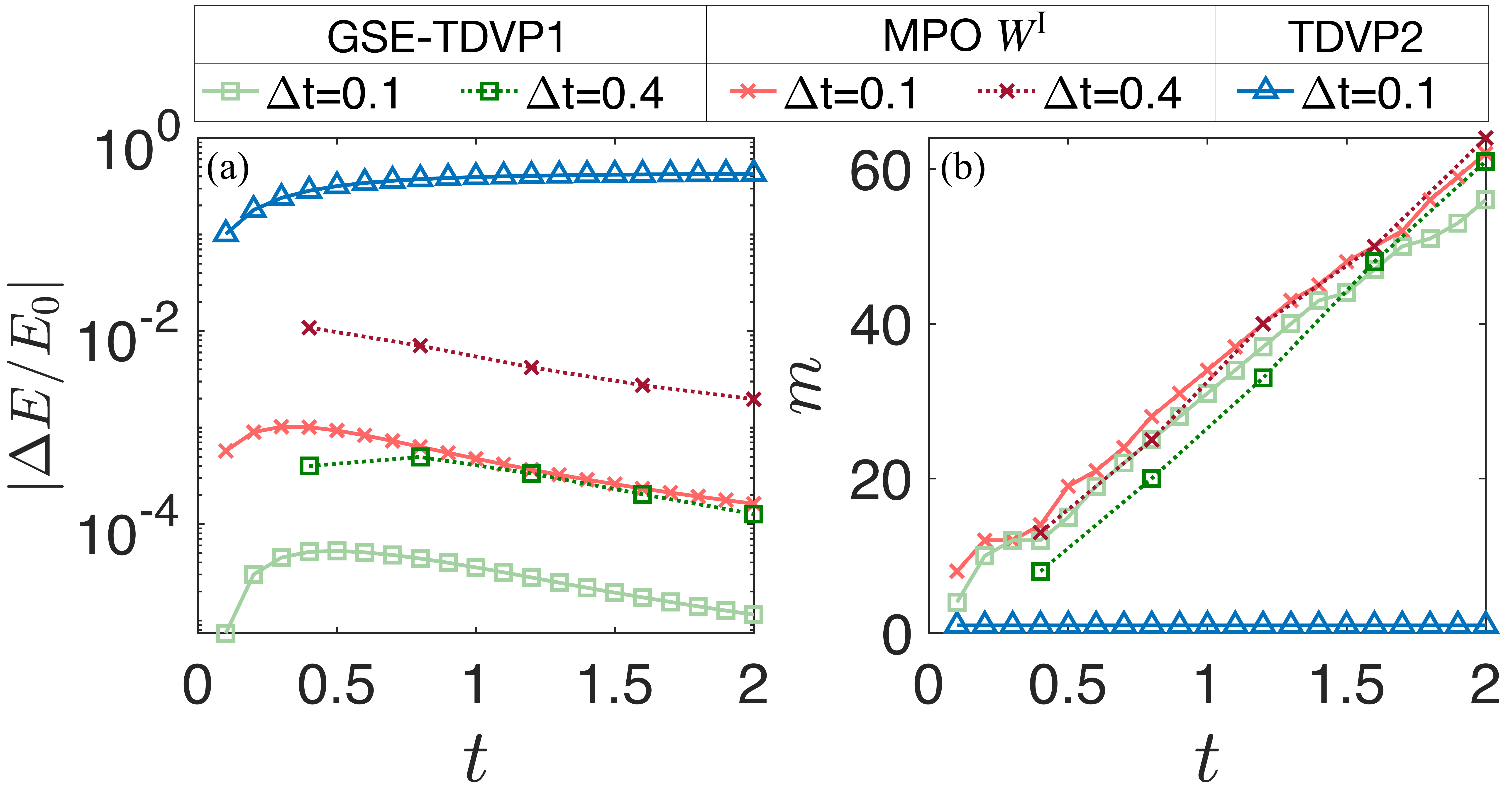}
\caption{\label{fig:6} Benchmark results of the imaginary time evolution for the rung-decoupled Heisenberg ladder. In all methods, $\epsilon=10^{-10}$. For MPO $W^{\mathrm{I}}$, $\epsilon$ is the truncation error in applying the MPO. For GSE-TDVP1, we use the optimal settings $\mathrm{i}\tau=1/40$, $k=3$, $\epsilon_M=10^{-8}$, and $\epsilon$ is the truncation error in the follow-up TDVP1 sweep. (a) Absolute energy error $\Delta E= E-E_{\mathrm{exact}}$ scaled by the ground-state energy $E_0$, where the reference energy $E_{\mathrm{exact}}$ is obtained by doubling the energy of a single chain from TDVP2 with $\Delta t=0.01$. (b) Bond dimension growth versus time. All methods except TDVP2 (which stays in a product state) show similar bond dimension growth.} 
\end{figure}

We first consider imaginary time evolution of the rung-decoupled spin-1/2 Heisenberg ladder of Fig. \ref{fig:tba}, with Hamiltonian
\begin{equation}
\hat{H}=\sum_{r,\langle i,j\rangle}\hat{\bm{S}}_{r,i}\cdot\hat{\bm{S}}_{r,j},
\end{equation}
where $r\in\{1,2\}$ denotes which leg it is and $\langle i,j\rangle$ denotes the nearest-neighbor sites along each leg. 

We evolve in imaginary time starting from a Neel (product) state $|\psi(0)\rangle$. Measuring the energy versus time provides a reasonable test of the evolution. In Fig. \ref{fig:6} we show a comparison of our method with two other methods. The first one is an MPO method of Ref. \onlinecite{PhysRevB.91.165112}, specifically the $W^{\mathrm{I}}$ method, where complex time steps have been used to make the error second-order. The second is the conventional TDVP2. In our method, we use $\epsilon_K=10^{-12}$ when applying $1-\mathrm{i}\tau\hat{H}$ with the density matrix approach\footnote{The variational approach faces the same local minimum problem as DMRG does, which can be solved by ways suggested before.}. We show results for $\mathrm{i}\tau=1/40$, $k=3$, and $\epsilon_M=10^{-8}$, which turns out to be near optimal parameter settings in this case. We found that higher order $k$ or smaller truncation $\epsilon_M$ in the subspace expansion do not improve the accuracy (not shown). While TDVP2 fails as expected, GSE-TDVP1 (with $\epsilon=10^{-10}$) has an accuracy 10 times better than the MPO $W^{\mathrm{I}}$ method with a comparable bond dimension growth. We also show GSE results with a larger time step $\Delta t = 0.4$, which exhibit a still reasonable error of $10^{-3}$. Not shown are results for $\epsilon=10^{-12}$, which are slightly more accurate than $\epsilon=10^{-10}$, but which also exhibit faster bond dimension growth. 

\subsection{Real time evolution}
\begin{figure}
\includegraphics[width=8.6cm]{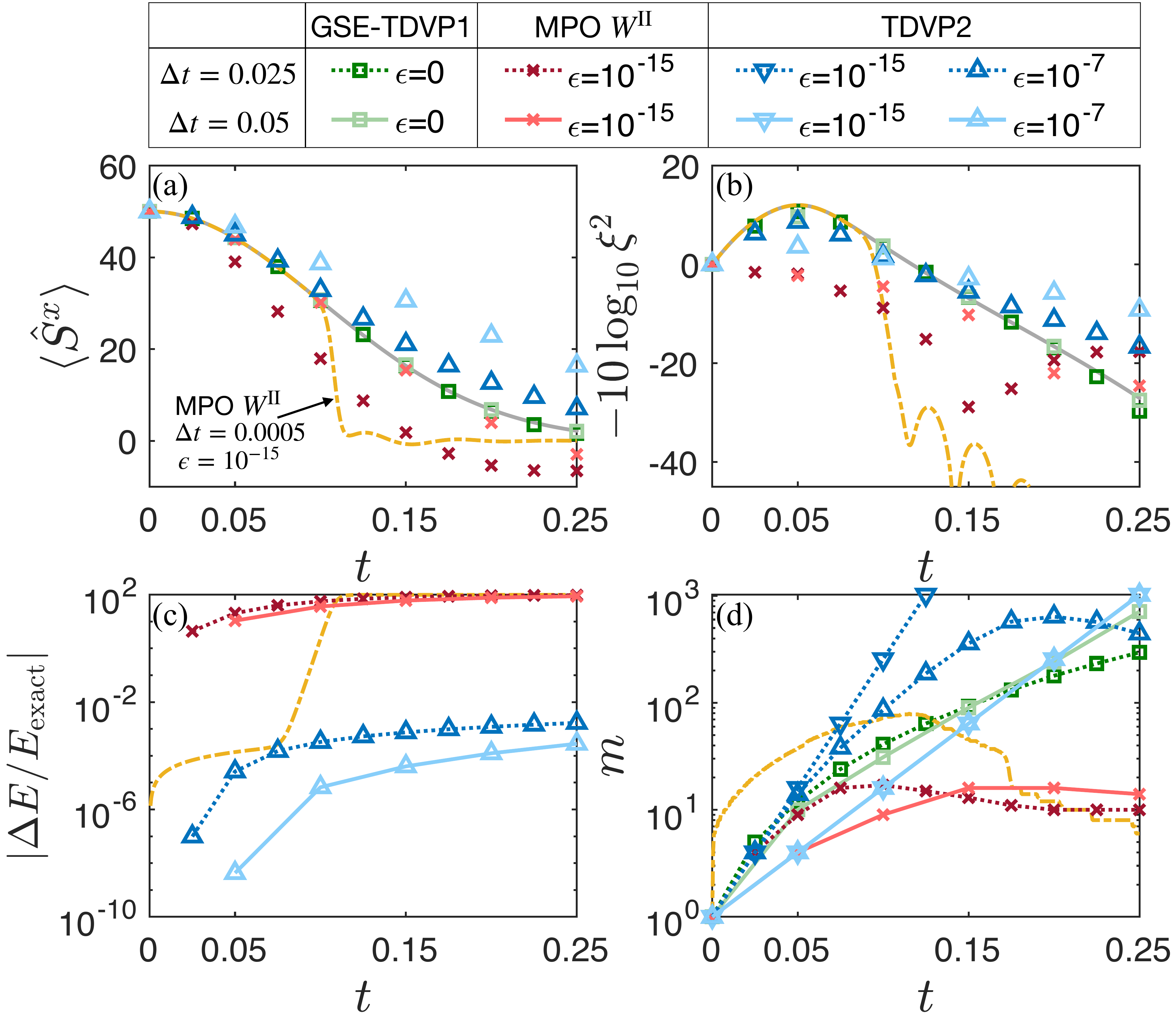}
\caption{\label{fig:7} Benchmark results of the real time evolution for the OAT model for a variety of methods, versus exact results (solid grey lines). (a) shows $\langle \hat{S}^x(t)\rangle$. GSE-TDVP1 is the most accurate method. The apparent accuracy of one of the MPO $W^{\mathrm{II}}$ calculations appears to be accidental. TDVP2 becomes more accurate with a smaller time step. (b) shows the spin squeezing parameter. Again, GSE-TDVP1 is the most accurate. (c) shows relative errors in the energy. Curves for GSE-TDVP1 and for TDVP2 with $\epsilon=10^{-15}$ are all below $10^{-10}$ and are not shown. Errors for the MPO $W^{\mathrm{II}}$ are particularly large. (d) shows the bond dimension growth. The smallest bond dimensions come from the MPO $W^{\mathrm{II}}$ methods but this is due to their large errors. TDVP2 with smaller time step has particularly large bond dimension growth. Note that in both (a) and (b) the TDVP2 data points for $\epsilon=10^{-15}$ (not shown) and $\epsilon=10^{-7}$ coincide.}
\end{figure}

The one-axis twisting (OAT) model\cite{PhysRevA.47.5138} has been widely studied for the use of quantum metrology\cite{MA201189}. This model has infinite-range interactions, making it a challenge for MPS methods, but it also has an exact solution. We study the real time evolution of $N=100$ spin-1/2's with the Hamiltonian
\begin{equation}
\hat{H}=\chi (\hat{S}^z)^2,
\end{equation}
where $\hat{S}^z=\sum_{i=1}^N\hat{S}^z_i$ and we set the energy scale $\chi=1$. The MPO representation of the Hamiltonian is rather simple, with a bond dimension $w=3$. We take as initial state all spins polarized in the $+x$ direction. During the time evolution the spin is squeezed\cite{perlin2020spin}. The exact solution for the $x$-moment is
\begin{equation}
\langle \hat{S}^x(t)\rangle=\frac{N}{2}\cos^{N-1}(t),
\end{equation}
with $\langle \hat{S}^y(t)\rangle=\langle \hat{S}^z(t)\rangle=0$.

An important property is the spin squeezing parameter, defined\cite{PhysRevA.46.R6797} as
\begin{equation}
\xi^2=N\min_{\bm{\mathrm{n}}_{\bot}}\frac{\langle(\bm{\hat{S}}\cdot\bm{\mathrm{n}}_{\bot})^2\rangle-\langle\bm{\hat{S}}\cdot\bm{\mathrm{n}}_{\bot}\rangle^2}{\langle\bm{\hat{S}}\rangle^2},
\end{equation}
where $\hat{S}^\mu=\sum_i\hat{S}_i^\mu$ and $\bm{\mathrm{n}}_\bot$ is a unit vector perpendicular to $\langle\bm{\hat{S}}\rangle=\langle\hat{S}^x\rangle\bm{\mathrm{n}}_x$. Minimizing over $\bm{\mathrm{n}}_\bot$, $\xi^2$ can be expressed in terms of correlation functions
\begin{equation}
\frac{\xi^2}{N}=\frac{\sigma_{yy}^2+\sigma_{zz}^2-\sqrt{(\sigma_{yy}^2-\sigma_{zz}^2)^2+(\sigma_{yz}^2+\sigma_{zy}^2)^2}}{2\langle\hat{S}^x\rangle^2},
\end{equation}
where $\sigma_{xy}^2\equiv\langle \hat{S}^x(t) \hat{S}^y(t)\rangle$, etc. The optimal spin squeezing $\xi^2_{\mathrm{opt}}$ is expected to appear at $t_{\mathrm{opt}}=12^{\frac{1}{6}}(N/2)^{-\frac{2}{3}}/2 \approx 0.05$. We continue the time evolution to $t=0.25$, or about $5t_{\mathrm{opt}}$.

In Fig. \ref{fig:7}, we compare our method with TDVP2 and MPO $W^{\mathrm{II}}$ (which is expected to work better than $W^{\mathrm{I}}$ here). To preserve exact unitarity, for GSE-TDVP1 we set $\epsilon=0$, which turns out to have a minimal extra cost in bond dimension, with the bond dimension already controlled by $\epsilon_K$ and $\epsilon_M$. Since $w$ is small, we use the density matrix approach to apply the MPO. For time step $\Delta t=0.025$, we find optimal parameters $\tau=\Delta t$, $k=3$, $\epsilon_K=10^{-4}$, and $\epsilon_M=10^{-4}$, which balance cost and accuracy. For time step $\Delta t=0.05$, we use $\tau=\Delta t$, $k=5$, $\epsilon_K=10^{-4}$, and $\epsilon_M=10^{-8}$. Our method is the most accurate and preserves unitarity exactly, while also having slower bond dimension growth than TDVP2. For MPO $W^{\mathrm{II}}$, the conservation of energy is very poor and the overall shape of $\xi^2$ is wrong. Reducing the time step size to $\Delta t=0.0005$ (yellow curves in Fig. \ref{fig:7}) helps initially, but the evolution soon becomes unstable after $t_{\mathrm{opt}}\approx 0.05$.

\section{Conclusions}
In this paper, we discussed how TDVP can fail in simple situations, and present a new algorithm, a modification of TDVP, which appears to work in all situations, including long-range interactions. The key modification is the enlargement of the tangent space before each time step using global Krylov vectors. The enlarged space introduces the degrees of freedom for the correct time evolution, allowing us to combine it with the single-site TDVP method. For real-time evolution we can maintain exact unitarity, even as the bond dimension is allowed to grow. (The two-site TDVP method can be viewed as a simpler attempt to enlarge the tangent space, which may not work well with non-nearest-neighbor interactions.) The new method does not require the time step to be made particularly small, and works correctly for evolution starting with a product state. Finally, our method has excellent efficiency, with a calculation-time cost between that of one- and two-site conventional TDVP. We expect it to be a valuable tool for out-of-equilibrium dynamics and finite-temperature simulations in systems with long-range interactions and in two dimensional systems.

\begin{acknowledgments}
We thank Matthew Fishman, Alexander Wietek, Chia-Min Chong, Sean R. Muleady, Edwin M. Stoudenmire, and David J. Luitz for helpful discussions. The algorithms are implemented using the ITensor\cite{itensor} library and the codes are available under the ITensor/TDVP repository. This work is funded by NSF through Grant DMR-1812558.
\end{acknowledgments}

\bibliography{mybibtex}

\end{document}